# From Grid Middleware to a Grid Operating System

Arshad Ali[1], Richard McClatchey[2], Ashiq Anjum[1,2], Irfan Habib[1], Kamran Soomro[1],
Mohammed Asif[1], Ali Adil[1], Athar Mohsin[1]

[1]National University of Sciences and Technology, Rawalpindi, Pakistan
Email:{Arshad.Ali,Irfan.Habib,Kamran.Soomro,asif.khan,ali.adil, athar.mohsin}@niit.edu.pk
[2]CCS Research Centre, Univ. of West of England, Frenchay, Bristol BS16
Email:{Richard.McClatchey,Ashiq.Anjum}@cern.ch

## Abstract

*Grid computing has made substantial advances during the last decade. Grid middleware such as Globus has contributed greatly in making this possible. There are, however, significant barriers to the adoption of Grid computing in other fields, most notably day-to-day user computing environments. We will demonstrate in this paper that this is primarily due to the limitations of the existing Grid middleware which does not take into account the needs of everyday scientific and business users. In this paper we will formally advocate a Grid Operating System and propose an architecture to migrate Grid computing into a Grid operating system which we believe would help remove most of the technical barriers to the adoption of Grid computing and make it relevant to the day-to-day user. We believe this proposed transition to a Grid operating system will drive more pervasive Grid computing research and application development and deployment in future.*

## 1. Introduction

In the last decade Grid computing has made great stride, and virtually revolutionized high performance and high throughput computing for use in scientific applications. Grid middleware such as UNICORE [1], Globus [2] and others has significantly contributed to making this possible, however limitations have arisen which restrict the adoption of Grid computing in other fields. Some of the most profound issues are: the strong focus of Grid computing research on the application layer [3], the overhead of managing and maintaining a Grid middleware and the limited support for a number of applications, all of which are irrelevant to the needs of common users. Other drawbacks of Grid middleware are its primitive resource management capabilities and the rather inflexible network topologies which are created by them. This paper will follow the following format: first we will review current popular approaches to building Grids, and analyze some limitations which are directly linked to the Grid middleware, then we will try to formally define the term Grid Operating System (Grid OS), afterwards we will look at a Grid OS which will address the limitations identified in the paper, and present an anatomy of it.

## 2. The Current Grid Climate

The most popular ways of building Grids are the so-called "cluster of cluster" approaches, where geographically separated clusters are linked via high-speed networks and a software infrastructure is installed to make them interoperate with each other.

Some of the largest Grids in the world are of this type [4] and [5]. The "software infrastructure" mentioned above, is generically termed "Grid middleware", some popular Grid middleware being: Globus, UNICORE and Legion [6]. Other approaches to Grid computing include the approach used by Google Compute [7] and projects based on BOINC [8] where a small program is installed in host computers and single or multiple servers coordinate the execution and retrieve results of some computation from potentially thousands of hosts.

Many technologies such as the World Wide Web (WWW) have been spun out from research projects which were targeted to facilitate scientific research. The WWW was an instant success both with the scientists and the general public, however although the Grid has been rapidly adopted, it has not, as yet, been accepted by common users. Science research labs such as CERN have used cluster computing during a large part of the 1990s to meet their computation needs.

Since Grid middleware for eScience has mainly focused on combining the power of computing clusters distributed worldwide many of the shortcomings in cluster computing have been brought into the Grid arena. Indeed, some of the largest Grids in the world (e.g. [4] and [5] ) are made from constituent clusters.

Day-to-day computer users and small to medium sized organizations often do not use clusters, and thus for them setting up Grids using the existing middleware is complex. Furthermore, Grid enabling their applications is nearly impossible, as they are not easily supported, and this poses a massive barrier to the pervasive adoption of Grid computing by these communities. Grid Middleware is also complex to setup and necessitates a steep learning curve. The difficulty in setting up Grid middleware is thus restricting the uptake of Grid computing in non-research environments (see [9]). Grid Portals have drastically reduced the complexity for the user to use Grids; however, the Grid portal itself has significant drawbacks as has been outlined by J. Marsh et al [10].

Current Grid infrastructures support only a few types of applications [11]; all supported job types are similar in that their input is predetermined, and applications have to be specially designed to be executed over a Grid. Grid Middleware relies on cluster execution services such to process applications and since clusters were never designed to support interactive applications, current Grid infrastructures also cannot support them. These applications are mostly used on PCs, however existing Grid middleware also does not offer any 'out-of-the-box' capability for synchronizing distributed processes. Projects such as MPICH-G/2[12], GridMPI[13] go a long way in addressing this problem; however applications have to be coded to the MPI standard[14].

## 3. Existing Grid Middleware and the Case for Migration

Existing Grid middleware supports primitive forms of resource management: Grid middleware such as Globus mostly facilitates Grids which are aggregations of clusters. The middleware does not enforce any rules on how many resources a job can consume because there is no way for the administrator to specify how much or to what degree he wants to share his resources on a Grid. This is a direct consequence of the Grid middleware being built on cluster execution services. Clusters have been deployed in dedicated environments and thus restricting resource usage to certain limits was never a realistic option. Various schemes based on usage policies, resource reservation and quotas have been implemented to address this restriction, however each are forms of workarounds and do not provide effective resource management. On the other hand Grid computing projects such as SETI@HOME[15] restrict themselves to harvesting only idle cycles from the host computers. In day-to-day user-oriented Grids, we need fine grained resource management, where users can monitor, and set limits on resource usage, anytime for anyone.

Existing Grid middleware is used to establish Grids which are of a hierarchical nature, and mostly follow the client-server paradigm. This paradigm works well in organizations such as research labs, or enterprises, which are hierarchical in nature, although such Grids can be linked to act as peers with collaborating research institutions or enterprises. Having many nodes on the Grid equates to larger number of points of failure, consequently in a hierarchical topology if the top tier fails, large portions of the Grid can fail, drastically reducing the overall Quality of Service (QoS). To minimize the impact of failures, to enable broader collaboration and to anticipate the changing and dynamic nature of Grids, a flatter approach to Grid computing should be considered, since hierarchical Grids at the user level would not be sustainable, due to an implicit lack of trust between the participating entities, and to potential failures of nodes. The Peer-to-Peer (P2P) paradigm has gained credence as a possible solution to this problem [16]; however there is as yet no facility to create P2P Grids with existing Grid middleware.

Most of the issues highlighted in the previous sections, can be resolved effectively by migrating many of the functions for which Grid middleware has traditionally been responsible to the operating system level. Migrating Grid computing to the operating system level would make it transparent and invisible to the user and would remove the obstacles related to the installation and maintenance of Grid middleware.

Grid enabling process management in the operating system yields a completely new paradigm for developing applications, which can remove many of the barriers related to re-engineering of application for execution on Grids and can provide universal support for many more types of applications; support for legacy desktop applications becoming a possibility. Resource management which has been carried out traditionally at the application level in Grid middleware would be done more effectively at the operating system level, since operating systems are basically resource management systems. Migrating Grid computing to the operating system, would remove any differences between non-Grid computing systems and Grid computing enabled systems, thus making Grid computing ubiquitous. Past precedents such as NFS and TCP/IP are solutions to common large problems, however, they are very simple to use and transparent to the user, because they have been integrated into modern operating systems. On similar lines we believe that by incorporating Grid computing features into modern operating systems a new paradigm of desktop computing can be encouraged. In this paper we discuss the high level architecture and features of such a Grid enabled operating system.

## 4. Towards a Grid Operating System

Projects such as GridOS[17] define a Grid OS to be an operating system with complementary support for Grid middleware. Projects such as Legion and WebOS[18] are aimed at the development of a middleware infrastructure for Grid computing. In this paper we refer to a Grid OS as a completely integrated operating system which integrates major Grid computing components into the machine operating system. We define a Grid OS to be an operating system which transparently enables a user to peruse discovered distributed resources, to share his resources in a P2P fashion, to launch and to migrate tasks on them and to enable the control and monitoring of executed processes. Ideally this would be a plug and play environment which autonomously discovers and makes network-agnostic calls to the Peers with built-in functions to transparently synchronize the local-to-global mapping of the resources. We envisage a Grid computing approach, where a single node in the Grid is not a cluster, but a single PC.and the following characteristics of a true Grid OS:

• The ability to transparently discover distributed resources and enable "Plug and Play" Grid computing. We propose a P2P approach to resource discovery (as described in section 4.4). This feature will give the users a system which requires minimum effort and configurations in constructing and joining Grids.

• The ability to transparently Grid-enable desktop applications. Grid-enabling interactive applications will be one of the major aims of the Grid OS. Local jobs will be synchronized with the global resources. We propose a new methodology in Grid enabling applications and discuss some issues in section 4.2 including the provision of high level abstractions to easily create Grid-enabled interactive applications.

• The ability to provide sufficient security to ensure peace of mind for both resource providers and consumers. Security is crucial for the success of any distributed system. We discuss end-to-end security issues with such a system in section 4.5

• The ability to provide support for legacy Grids, such as interoperability with existing Grid middleware. To implement support for existing middleware infrastructures the operating system would require embedded interfaces enabling interoperability.

• The ability to be network aware. Networking technologies are predicted to double in bandwidth every nine months; however application requirements are also increasing exponentially. It is becoming increasingly important, to manage the network as a resource, and designing the Grid OS to be network aware, is an important goal.

Assume a scenario where a user runs an interactive program on his PC. In contemporary operating systems such as MS-Windows and GNU/Linux, all user interaction, input/output and processing is handled at the user's PC. In existing Grid middleware, the user would have to login to the Grid portal, create a job description, and upload the file and associated data, which itself might not be available at that time for Grid scheduling. In a Grid OS the user would feel just as if he is using a conventional operating system; however the kernel level takes care of all aspects of: a) Resource brokering and Grid-wide scheduling, discovery of resources; b) Mapping of "parts" of local processes to global processes, and process status tracking and c) Security and confidentiality of data.

All of these aspects would be handled in a P2P manner between machines transparently by the user, thereby leading to a huge increase in the response time of the user application and decreasing the complexity of Grid computing software dramatically. Using existing Grid middleware this would not be possible. On the other hand a Grid OS should not be built without considering interoperability support for existing Grid middleware. The existing Grid computing infrastructure can itself benefit from the introduction of the desktop to the infrastructure. There were estimated to be close to 590 million PCs in the world in 2003, and by now this number is much greater; each PC nowadays is equally capable if not more powerful then the supercomputers of the 1980s. It should therefore be possible for any organization including governments to create adhoc Grids, and have desktop users subscribe to them and share their resources, according to some sharing policy defined by the user. This has great potential for organizations which do not have the resources to purchase and setup clusters or supercomputers; with Grid enabled desktops they can utilize existing commodity PCs. Therefore a Grid OS should focus on achieving a dual objective: bringing

Grid computing to the desktop, and bringing the desktop to the Grid.

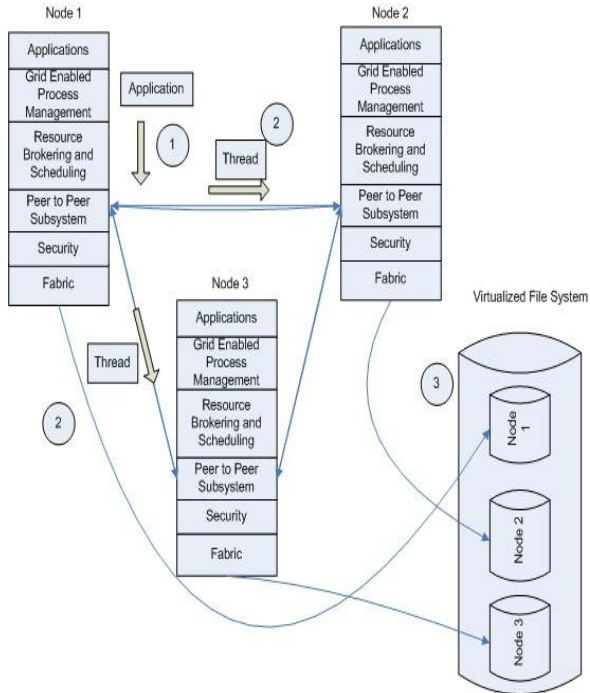

Figure 1. Proposed High-level Architecture of the Operating System.

## 5. A Proposed Architecture for a Grid OS

Figure 1 shows a sample architectural interaction between Grid OS components when the user submits a job. The user executes an application and the operating system's Grid Enabled Process Management system automatically detects that this is a Grid enabled job and creates its execution threads, which it then forwards to the Resource Brokering and Scheduling engine where the threads are moved for processing to nodes that are more powerful than the user's own node. The communication between the nodes takes place via the P2P subsystem, which is also responsible for the discovery of resources. A Security layer shields the user's primary resource such as data stores, memory etc. All these resources are collectively termed as "Fabric" in the diagram. The file system of the individual operating system would be a server-less P2P file system, where from any one node the user is capable of viewing universally the entire shared data space. In the following sections we will discuss all the major portions of this operating system.

## 6. Salient Features of the Grid Operating System Approach

### 6.1 P2P Resource Broker and Scheduler

A Resource Broker is a central component in any Grid computing environment. The purpose of a Grid Resource Broker is to dynamically find, identify, characterize, evaluate, select, allocate and coordinate resources with different characteristics most suitable to the user's application. Most existing Resource Brokers require significant user intervention and involvement to operate and they are designed for batch applications. Thus these Brokers are not feasible for the end-user who is concerned with ease of use, and high response and minimum turnaround time of jobs, which are not supported by existing Grid resource brokers. Present desktop operating systems take brokering and scheduling decisions by considering only local (rather than global) resources.

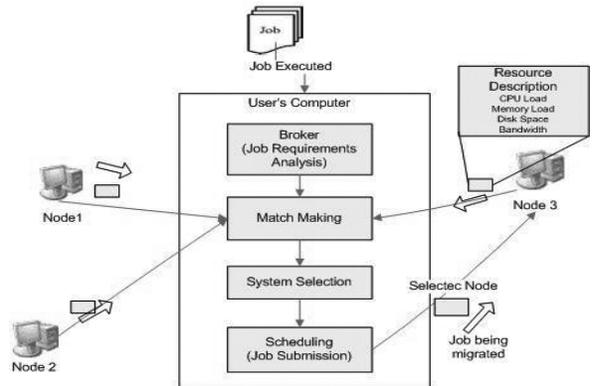

Figure 2. Architectural Overview of the Resource broker and scheduler

The Grid computing environment is a dynamic environment where status and load on resources are subject to change. Hence in any such environment it is very complex for the Broker to predict the performance and efficiency of the application on any specific resource. This problem is being addressed by current Grid middleware through policy based scheduling, policy negotiation and advance resource reservation schemes. In this scenario the Broker has some kind of exclusive control over system resources in order to improve its performance and decision making ability. For the Grid OS we envisage a brokering and scheduling engine inside the operating system kernel which takes into consideration entire pools of resources available across the Grid. We propose a P2P resource broker framework in which everything from matchmaking of requirements and available resources, down to the scheduling is carried out cooperatively with Peers, thus enabling compute-intensive and memory-intensive applications, making best use of the Grid resources in order to achieve high throughput. In our proposed framework, shown diagrammatically in Figure 2, each resource in the system will advertise its most recent status dynamically. The task of the Broker is to collect this information and select the optimum machine among the eligible machines based on the job's characteristics and the requirements submitted by the user. The job requirements, that are the computational demands of the application, will be specified by the developers of the application in the form of attached metadata. Eligible machines are those machines which fulfil the user's minimum requirements. Integrating the Grid Resource Broker into the operating system kernel is one of the first steps towards providing transparent access to Grid resources for users' applications.

## 6.2. Grid Enabling Interactive Applications and File Systems

Currently the avenues of Grid enabling interactive applications are somewhat limited, and significant changes need to be made to such applications in order to Grid-enable them. Grid-enabling interactive applications could remove some of the usability related obstacles to the more widespread adoption of Grid computing and could limit the time to create Grid enabled applications. Users would then not need to specify workflow plans [19] to dispatch applications to the Grid, rather this can be handled transparently by the operating system. We believe our approach is simple, yet powerful: nearly all interactive applications are built around the multithreaded paradigm, where separate threads of execution, sometimes interdependent, handle the execution of the application. The kernel of the Grid OS will have support for thread migration, an extension of the concept of process migration [20], popular in cluster middleware. In our system, the kernel would be capable of migrating a single thread, to another system, its broker finding it to be more powerful.. However migrating a thread in a Grid environment, where there is no shared memory, is not without issues, as identified by [21]. Thus to facilitate thread migration either the memory needs to be virtualized or the entire address space needs to be moved along with the thread. There are still many unresolved research issues related to implementing Grid wide shared memory, as has been explored by [22], thus our system will be a non-shared memory system.

When we talk about Grid-enabling interactive applications, we must also consider the Grid-enabling of the interaction between the threads of execution in a process. In the existing Grid infrastructure, inter-process communication across the Grid is non-existent, since applications are tailored to be batch-oriented, and execute independently. Communication between threads of the same process usually entails a procedure call, along with some means of identifying the particular thread with which to communicate. To Grid-enable such applications, the operating system should be able to intercept the communication, to determine the destination and to pass on the procedure call to the appropriate node for it to be executed there.

We plan to do this by having a service in the operating system that monitors all procedure calls by a thread scheduled to it. It would be able to determine whether the procedure called is a member of a remote thread. The information of which thread is scheduled to which node would be available to the operating system. Once the destination of a particular call is determined, the call could be forwarded to the concerned node to be executed there. Any data associated with the call must also be transferred to the other system. One issue in this approach is that of network latency in Grid environments. This approach requires significant amounts of data transfer. Although current network speed is much less then processor-memory speed, advances in networking technologies will eventually reduce the communication latencies involved. Another issue is that the Thread IDs must be unique across space and time i.e. they must be unique within the context of the whole Grid.

The challenge for Grid-enabling desktop file systems in our project will be to unify all shared storage devices, into a single logical namespace and to provide a universal view to each participating Peer. Security is one essential concern for us. There are many Grid-specific file system technologies [23] and [24] and others, most of which have been designed to be used in hierarchical environments. However projects such as [25] are P2P, serverless file systems, designed to be used in environments where there is incomplete trust, and to implement advanced security features. These projects are very relevant to our work, and could be re-used in our Grid Operating System.

## 6.3. Peer-to-Peer (P2P) Topology-Aware Discovery

In a Grid Operating System, each peer will most probably be either a Desktop PC or a Laptop PC. It is accepted that they will be physically distant from each other. There should be some mechanism in which each peer shall prioritize the other peers with the highest throughput of information dissemination. Another problem is that single peers cannot handle or store the information regarding each peer in a system. In order to provide scalability and to make the information dissemination more efficient, we propose a P2P topology-aware discovery service, in which the peers will be aware of all the other peers based on the available bandwidth from each peer. Resource discovery is one of the cornerstones of any Grid system. We advocate a network topology-aware resource discovery, which aims at making the Grid OS itself network aware. There are many Grid Resource discovery mechanisms [26, 27, 28, and 29] but none of them is network topology-aware. Topology-aware discovery services will enable the Grid OS to peruse the network infrastructure more optimally. In a Grid OS, some Peers will have variable network connectivity. The Peers that are closest to each other in our case are those which have a low round trip time (RTT) between them. Low RTT brings high bandwidth and hence faster communication. The proposed topology for our Grid OS should lead to fast and reliable communication between Peers. The goal of the proposed service is that Peers should self-organize themselves into sub-Grids, in which all peers are nearest neighbours to each other. All sub Grids and their peers will be members of the RootGrid.

To demonstrate our approach, consider that we have six Peers and we need to find the nearest Peer among these six Peers with respect to our site. These Peers are named from 1-6. In order to find the nearest Peer, we have two options: either to calculate only the RTT from these Peers or to calculate available bandwidth to these Peers. In our case, we need more than just the RTT as the factor to find the nearest node. The other performance parameters that we propose to use include Packet Loss and Available bandwidth. As described in [30], we can estimate the upper bound of available bandwidth as a function of RTT, Packet Loss and MSS :

$$BW < (MSS/RTT)*(1/sqrt (Packet Loss))$$

As we see in Figure 3, R6<R4<R2<R5<R3<R1; this shows that Peer 6 is the nearest node and then Peer 4 and so on. But in Figure 4, we see that B4>B6>B1>B5>B2>B3. It means that Peer 4 has a better network performance as compared to other Peers from our site. The reason behind these different results is that bandwidth doesn't depend only on RTT, rather it also depend upon Packet Loss. In order to define nearest Peer, let us say we have N destination Peers, and we have to find the nearest Peer among them. B1, B2, B3,…, BN is the estimated available bandwidth from source to N destinations.

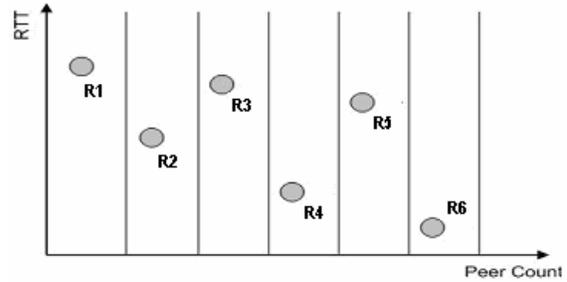

Figure 3. Round Trip Time from source 1-6 to destinations

Using this notation, we define the nearest Peer as:

NP = { Bk | max (Bi) , 0<i<N }

In this way, the Peer with maximum available bandwidth from source is our NP.

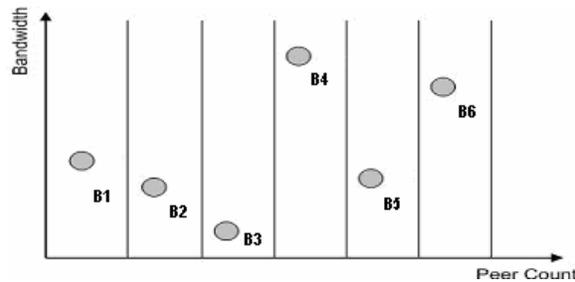

Figure 4. Available Bandwidth from source to 1-6 destinations

In our approach, each peer will have a unique ID, which will be assigned at the time of its joining the Grid. After joining, it will check for the existence of RootGrid. If RootGrid does not exist, it means this is the first peer joining the system. That peer will then create RootGrid and will join it. If RootGrid exists then the peer will automatically join RootGrid and will search for the nearest subGrid and will join that subGrid. When the new SubGrid is created, due to its limitation factor "lim" in each sub Grid, all the other peers in a system will be notified for the new sub Grid. Each peer will then update its nearest SubGrid table and may change or remain in the Grid which is nearest to that Peer. This algorithm will setup the following topology, as shown in figure 5. Figure 5 shows three Sub Grids and the RootGrid is the sum of all these Grids.

In Figure 5 the system is shown based on a decentralized architecture. The basic concept of this topology architecture is that each Grid will have its Master and Slaves. There is no restriction on slaves that communicate with each other directly. The only responsibility of the Master is to propagate the Grid's information to other known Grids through their

Masters. After getting new information the Master propagates new information to all slaves.

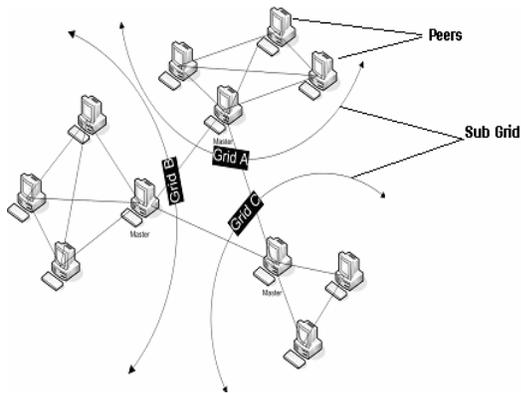

Figure 5. Topological Structure

When the Master goes down, one of its slaves is nominated to become Master, based on performance. The basic role of the Master-Slave relationship is discovery. The Slave registers its network, application and other relevant resources with itself and the Master. The Master propagates that information to all other Masters and they consequently inform their own slaves about the new resources added to the system.

### 6.4 End-to-End Security Framework

We believe the single most important determinant in the success of the Grid OS project will be security. In a Grid OS, resource providers need to be assured that no malicious programs will execute on their system and resource consumers must be assured that their data is safe in a Grid environment. Grid middleware provides no security in terms of preventing execution of malicious programs. However if the Grid is to be used in everyday user environments, a more comprehensive security framework must be designed. For a Grid OS we envisage an end-to-end security framework, which not only ensures users privacy and confidentiality of their data but also allows the user to set limits on resource usage, and to audit the same. A security framework for the Grid OS will encompass three components: Resource Provider security, Resource Consumer security and Transport security. Computer resources, such as processor cycles, memory and storage resources are provider by resource providers to foreign users.

Without a credible security framework for resource providers to do so, no Grid system can function. Resource providers not only need surety that their resources will not be misused or worse perform a malicious act on their own computer system; they also need mechanisms to control their level of commitment to the Grid. The providers would be given tools with which they can configure resource usage levels and monitor them, and be notified of any violations. Some of the techniques developed in the past decade such as mandatory access controls within the kernel operating system are important features to control potential abuse. The users whose jobs are executing in the Grid need to be assured that their data is safe, and that their execution's result cannot be tampered with by any external entity. Any Grid OS must explore the issue of keeping Grid scheduled data safe, in keeping the address space of the exported job in an area of the resource provider's memory which the user himself cannot read or write, however the resource providers will have facilities to control the size of this area. Another possible solution is to run the exported program in a sandbox, and completely separate it from the resource provider's own jobs. The Grid OS must ensure that data being transmitted between the nodes must be secure and it must also ensure its integrity when it reaches its destination. In a Grid OS there will be no central certification authority to handle authorization certificates. Rather each Peer must negotiate resource usage individually. However individual access negotiations are not scalable, and thus the user of proxy delegation is popular in Grids which leads to big performance boosts. A major issue in our project is how to handle privilege delegations in a Peer to Peer manner.

## 7. Conclusions and Future Work

As discussed in this paper Grid computing has been targeted to some segments of computer users, because of which various limitations have been built into existing Grid middleware. No conscious effort has been made to remove the barriers to adoption of Grid computing in user centric computing environments. In this paper we analyzed the most profound problems which are limiting the adoption of Grid computing in those fields. We believe an operating systems approach overcomes the problems highlighted in the paper, and takes the first step towards the creation of a pervasive framework for a Grid OS which would enable not only home and business users but also scientists, to do away with dedicated computing facilities for compute/data intensive application processing and to process their applications ron their own (or neighbouring) desktops. As a consequence hitherto unprecedented levels of collaboration can be achieved between users on the same computer resources. The Grid OS would also make fundamental contributions to existing Grid infrastructures e.g. the creation of enhanced P2P network topologies for Grids and enabling the execution of a multitude of Grid-enabled applications in addition to facilitating existing Grids to peruse millions of PCs located worldwide.

The ultimate aim of our project is to bring Grid Computing to the Desktop, and the Desktop to the Grid. This paper discussed some components which would bring Grid Computing to the Desktop, mainly focusing on the lowest level in the OS, the kernel. In future we will focus on the higher levels, primarily the interoperability of the Grid Operating System, with the existing prevalent web services based Grid middleware infrastructures In addition, we aim to provide a reference implementation of the operating system based on Linux which integrates the mentioned concepts and technologies described in this paper.